\documentclass[floats,aps,prb,showpacs,twocolumn,superscriptaddress]{revtex4}
\usepackage{graphicx}
\usepackage{amsmath,amssymb,latexsym,amsfonts,subfigure,bbm}

\def\Sb{{\mathbf S}}

\newcommand{\be}{\begin{equation}}
\newcommand{\ee}{\end{equation}}
\newcommand{\beqn}{\begin{eqnarray}}
\newcommand{\eeqn}{\end{eqnarray}}
\newcommand{\vare}{\varepsilon}

\newcommand{\bml}{\begin{multline}}
\newcommand{\eml}{\end{multline}}
\newcommand{\hath}{\hat{H}}

\newcommand{\hatsb}{\hat{\mathbf S}}
\newcommand{\phanda}{\phantom{\dagger}}

\newcommand{\Ah}{\hat{A}}
\newcommand{\Bh}{\hat{B}}
\newcommand{\Ch}{\hat{C}}
\newcommand{\Dh}{\hat{D}}
\newcommand{\Eh}{\hat{E}}

\newcommand{\hatp}{\hat{P}}

\newcommand{\deltab}{\bar{\delta}}
\newcommand{\rhob}{\bar{\rho}}

\begin{document}

\title{The strong-coupling limit of a Kondo spin coupled to a mesoscopic quantum dot: effective Hamiltonian in the presence of exchange correlations}

\author{Stefan Rotter}
\affiliation{Institute for Theoretical Physics, Vienna University of
Technology, A-1040, Vienna, Austria, EU}
\author{Y.~Alhassid}
\affiliation{Center for Theoretical Physics, Sloane Physics Laboratory, Yale  University, New Haven, CT 06520, USA}

\begin{abstract}
We consider a Kondo spin that is coupled antiferromagnetically to a
large chaotic quantum dot. Such a dot is described by the so-called universal Hamiltonian and its electrons are interacting via a
ferromagnetic exchange interaction. We derive an
effective Hamiltonian in the limit of strong Kondo coupling, where the screened Kondo spin effectively removes one electron from the dot. We find that the exchange coupling constant in this reduced dot (with one less electron) is renormalized and that new interaction terms appear beyond the conventional terms of the strong-coupling limit. The eigenenergies of this effective Hamiltonian are found to be in excellent agreement with exact numerical results of the original model in the limit of strong Kondo coupling.

\end{abstract}
\pacs{72.15.Qm, 72.10.Fk, 73.21.La, 73.23.Hk, }
\maketitle
\section{Introduction}
The Kondo resonance, which emerges when a localized impurity spin interacts antiferromagnetically with a delocalized electron gas,  has generated considerable interest over several decades by now.~\cite{kondo64,hewson93}  The observation that the Kondo resonance can be realized in the mesoscopic regime of quantum dots, in which many of the system parameters are experimentally tunable has led to much renewed interest over the last decade.\cite{goldhaber98,leos,cronenwett98,schmid98,simmel99,vdwiel00,craig04,jarillo,potok07,hubel,reyes09}
This experimental work has been accompanied by much theoretical progress on the mesoscopic aspects of the Kondo problem.\cite{glazman88,ng88,meir91,thimm99,simon01,pusti,simon02,cornaglia03,murthy05,kaul05,kaul06,sindel07,vitu,rotter08}

\begin{figure}[!b]
\includegraphics[angle=0,width=80mm,clip=true]{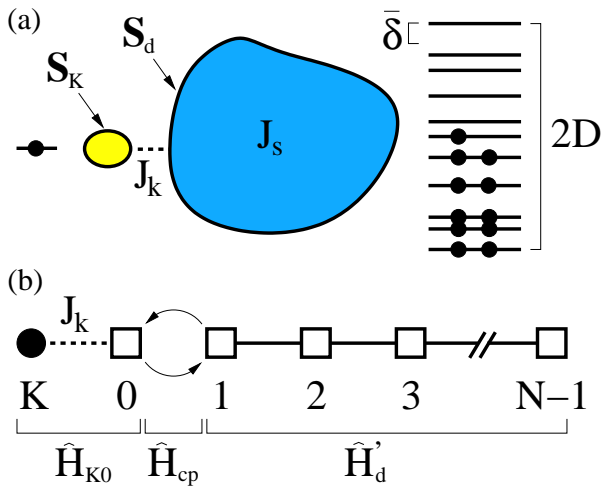}
\caption{(Color online) (a) Schematic illustration of the system under consideration: a small quantum dot (left) with spin
${\bf S}_K$ (Kondo spin) is coupled antiferromagnetically (coupling
constant $J_k$) to a large quantum dot (right) with spin ${\bf S}_d$. The large dot is described by the universal Hamiltonian, characterized by a ferromagnetic exchange interaction (coupling constant $J_s$). The $N$ single-particle energy levels in the large dot are distributed within a band of width $2D$ (half-filling). The
average single-particle level spacing is fixed and given by $\deltab$. (b) The large dot is represented in the site basis (squares), in which ${\bf S}_K$ couples only to site 0. In the strong-coupling limit, $J_k\to\infty$, it is useful to divide the Hamiltonian into three parts, $\hath_{K0},\,\hath_{cp},\,\hath'_{d}$, according to the sites involved [see Eq.~(\ref{eq:2bbb})].}
\label{fig:1}
\end{figure}

In the mesoscopic regime, the spin-1/2 Kondo impurity is typically represented by a small quantum dot with an odd number of electrons, while the delocalized electron gas is realized by electrons in leads or in a large quantum dot. In this work we focus on the latter case, assuming a small and a large quantum dots that are coupled antiferromegnetically as in Fig.~\ref{fig:1}a (see Ref.~\onlinecite{craig04} for an experimental realization of such a setup).

There are certain features that distinguishes the mesoscopic Kondo regime from the bulk limit. While the conventional Kondo theory assumes a continuum band of energy levels in the electron gas, the single-particle energy levels in the large quantum dot are discrete. The discreteness and the dot-specific realization of these energy levels become important when the Kondo temperature $T_K$, the characteristic energy scale of the correlated Kondo resonance, is of the same order of magnitude or smaller than the average level spacing $\bar{\delta}$.\cite{thimm99,simon02,murthy05,kaul06,rotter08}
In the conventional bulk Kondo model the electron-electron interactions in the electron gas are often neglected. However, for the present double-dot setup, electron-electron interactions in the large dot can play an important role.  In the following we assume the single-particle dynamics in the large quantum dot to be chaotic,\cite{jalabert,folk,chang,al00} in which case the dot
is described by the so-called ``universal Hamiltonian''.~\cite{uniham}
This Hamiltonian describes the low energy physics in a Thouless energy
interval around the dot's Fermi energy. For a semiconductor quantum
dot with a fixed number of electrons and in the limit of a large
Thouless conductance, the electron-electron interaction is dominated
by a ferromagnetic exchange interaction that is proportional to the
square of the total dot spin. Despite its conceptual simplicity, this universal Hamiltonian description was shown to yield a quantitative agreement\cite{rupp} with experimental results measuring the statistics of the Coulomb blockade peak heights and spacings.\cite{patel}

The effect of ferromagnetic exchange correlations
on the Kondo resonance was first addressed analytically in the bulk limit,\cite{larkin72} and, more recently, mean-field studies were carried out in the mesoscopic regime.\cite{murthy05}
In a recent work, we studied this problem numerically and provided analytical results for the weak and strong Kondo coupling limits.\cite{rotter08}  We found that for weak Kondo coupling, the Kondo spin acts like an external magnetic field, assisting the ferromagnetic polarization of electrons in the large dot. In the case of strong Kondo coupling, the Kondo spin effectively removes one of the electrons of the large dot. We showed that this ``reduced'' dot with one less level and one less electron can again be described by a universal Hamiltonian but with a renormalized exchange constant.

A central issue that was not addressed in our previous work concerns the nature of residual interactions in the reduced dot beyond the renormalization of its exchange interaction term.  From the work of Nozi\`eres\cite{nozi} we know that a {\it non-interacting} electron gas turns into a Fermi-liquid when strongly coupled to a Kondo spin. The dominant effective interaction between the quasi-particles in this Fermi liquid is a repulsive interaction between spins of opposite orientation that are in close proximity to the Kondo spin. In the present case, the finite exchange interaction in the large dot leads to new effective interaction terms in the strong-coupling
limit. To identify these new interaction terms, we follow a strategy that is similar to the one used by Nozi\`eres,\cite{nozi} i.e., we perform an explicit perturbative expansion of
the effective Hamiltonian of the reduced dot in the strong-coupling limit. In the presence of exchange interaction, this strong-coupling expansion is significantly more involved.
However, the resulting effective quasi-particle interaction contains
only a few new terms. The analytical expressions we derived for these effective exchange-like interactions are validated by a comparison with a full numerical diagonalization of the original two coupled dot model.

The outline of this paper is as follows: In section II we present the model of a spin-1/2 quantum dot that is Kondo-coupled to a large quantum dot (described by the universal Hamiltonian), and discuss the transformation from the single-particle orbital basis of the large dot to a chain site basis, commonly employed in the strong-coupling limit.  In Section III we discuss the limit of strong Kondo coupling and use a projection method to derive an effective Hamiltonian for the reduced large quantum dot with one less electron. In section IV we describe the evaluation of the eigenenergies of this effective Hamiltonian, and in section V we compare the results derived from this effective Hamiltonian with an exact numerical solution of the original model. In section VI we conclude with a summary and discussion.

\section{Model}

We consider a chaotic quantum dot that is coupled antiferromagnetically to a Kondo spin as realized,
e.g., by a small quantum dot with an odd number of electrons. A schematic illustration of such a system is shown in
Fig.~\ref{fig:1}a. In the following we will refer to the large quantum dot as the ``dot'' and to the small dot as the ``Kondo spin.''

\subsection{Hamiltonian}

In the limit of a large Thouless conductance, a quantum dot whose single-particle dynamics are chaotic is described by the universal Hamiltonian\cite{uniham}
\be\label{H-dot}
\hat{H}_d=\sum_{n=0}^{N-1}\sum_{\sigma} \vare^{o}_n\,
\hat{a}^\dagger_{n,\sigma}\hat{a}^{\phantom{\dagger}}_{n,\sigma}-J_s\,
\hat{\mathbf S}_d^2\,.
\ee
Here $\hat{a}^\dagger_{n,\sigma}$ creates an electron with spin
up/down ($\sigma=\pm1$) in level $n$ in an orbital single-particle
level with energy $\vare^{o}_n$. We assume $N$ spin-degenerate single-particle levels spanning a bandwidth of  $2D=(N-1)\times\deltab$ ($\deltab$ is the mean
level spacing). The second term on the r.h.s. of Eq.~(\ref{H-dot}) represents a ferromagnetic exchange interaction ($J_s >0$) where $\hat{\Sb}_d =
\frac{1}{2}\sum_{n \sigma\sigma'} \hat{a}^\dagger_{n\sigma}
\boldsymbol{\tau}_{\sigma\sigma'}\,\hat{a}_{n\sigma'}$
($\boldsymbol{\tau}$ are Pauli matrices) is the total spin of the dot. In Eq.~(\ref{H-dot}) we have ignored a constant charging energy term and a repulsive Cooper channel term.

The dot is coupled antiferromagnetically to a Kondo spin $\hat{\mathbf S}_K$ ($S_K=1/2$)\cite{murthy05}
\be\label{eq:1}
\hat{H}=\hat H_d +J_k\,\hat{\mathbf S}_K\cdot\hat{\mathbf s}_d(0)\,,
\ee
where $J_k$ ($J_k >0$) is the Kondo coupling constant and $\hat{\mathbf s}_d(0)$ is the spin density of the dot at the tunneling position ${\bf r}=0$.  The dot spin density at position ${\bf r}$ is given by
\be\label{eq:spindensity}
\hat{\mathbf s}_d({\mathbf r})= \frac{1}{2}\sum_{\sigma,\sigma'} \hat{\psi}^\dagger_\sigma({\mathbf r})
\boldsymbol{\tau}_{\sigma,\sigma'}
\hat{\psi}^{\phantom{\dagger}}_{\sigma'}({\mathbf r}) \,,
\ee
where $\hat{\psi}^\dagger_\sigma({\mathbf r})$ creates an electron
with spin $\sigma$ at position ${\bf r}$. In terms of the
single-particle orbital wave functions $\phi_n({\bf r})$, the field
operator is given by $\hat{\psi}^\dagger_\sigma({\mathbf
 r})=\sum_{n=0}^{N-1}\phi_{n}({\mathbf
 r})\,\hat{a}^\dagger_{n,\sigma}$ and the \emph{local} density of
states of the dot is given by\cite{thimm99,murthy05,kaul06}
$\rho(\varepsilon)=\sum^{N-1}_{n=0}|\phi_n(0)|^2
\delta(\varepsilon-\varepsilon^o_n)$, with an average value of
$\rhob\approx1/(N\deltab)$.

\subsection{Chain site basis}

The strong-coupling limit of the system in Fig.~\ref{fig:1}a is more
clearly described when the Hamiltonian
in Eq.~(\ref{eq:1}) is rewritten in a different basis, known as the
chain site basis.  This new basis is obtained by a unitary
transformation of the orbital basis\cite{hewson93}
\be\label{eq:utrans}
\hat{c}^\dagger_{\mu,\sigma}=\sum_{n=0}^{N-1}U_{\mu,n}\,
\hat{a}^\dagger_{n,\sigma} \,,
\ee
such that site $\mu=0$ corresponds to the tunneling position ${\bf
  r}=0$, and the new one-body site Hamiltonian of the dot is
tridiagonal, i.e., each site is coupled to its two nearest
neighbors. Such a transformation is constructed by choosing
$\hat{c}^\dagger_{\mu=0,\sigma}\equiv\hat{\psi}^\dagger_\sigma
({\mathbf r}=0)$ and carrying out a Gram-Schmidt orthogonalization
procedure.\cite{hewson93}

The chain site single-particle energies $\varepsilon^c_\mu$ are given by the diagonal elements of $\hat{H}_0= \sum_{n=0}^{N-1}\sum_{\sigma} \vare^{o}_n\,
\hat{a}^\dagger_{n,\sigma}\hat{a}^{\phantom{\dagger}}_{n,\sigma}$ when the latter is rewritten in the site basis. The off-diagonal
matrix elements $t_{\mu} \equiv t_{\mu,\mu + 1}$ and $t^*_{\mu} \equiv t_{\mu,\mu -1}$ describe the hopping amplitudes between neighboring sites. A spin $\hat{\mathbf s}_\mu$ can be associated with each site, where the spin of site $\mu=0$ is equal to the spin density at the tunneling position, i.e., $\hat{\mathbf s}_0\equiv\hat{\mathbf s}_d(0)$. In the site basis, the Kondo spin couples only to a single site $\mu=0$. The Hamiltonian in Eq.~(\ref{eq:1}) is now given by
\beqn\label{eq:2}
\hat{H}&=&\hat{H}_0 -\!J_s\,\hat{\mathbf S}_d^2\!
+\!J_k\,\hat{\mathbf S}_K\cdot\hat{\mathbf s}_0\,,
\eeqn
where the total spin of the dot is $\hat{\mathbf S}_d = \sum_{\mu=0}^{N-1} \hat{{\bf s}}_\mu$. Here $\hat H_0$ is the one-body Hamiltonian of the dot in the site basis
\be\label{eq:2a}
\hat{H}_0=\sum_{\mu=0}^{N-1}\sum_{\sigma} \varepsilon^{c}_\mu\,
\hat{c}^\dagger_{\mu\sigma}\hat{c}^{\phantom{\dagger}}_{\mu\sigma}
+ \hat H_{\rm hop} \;,
\ee
with $\hat H_{\rm hop}$ being the hopping Hamiltonian
\be\label{eq:2b}
\hat{H}_{\rm hop}=\sum_{\mu=0}^{N-2}\sum_{\sigma}\left(
t^{\phantom{\dagger}}_\mu \,
\hat{c}^\dagger_{\mu,\sigma}
\hat{c}^{\phantom{\dagger}}_{\mu+1,\sigma}\right.
+\,{\rm h.c.}\Big)\,.
\ee

The site basis formulation is particularly advantageous for the strong-coupling limit $J_k\rightarrow\infty$ when the site $\mu=0$ effectively decouples from the rest of the chain. Accordingly, we decompose the Hamiltonian in Eq.~(\ref{eq:2}) into three terms (see Fig.~\ref{fig:1}b for a schematic illustration)
\be\label{eq:2bbb}
\hat{H}=\hat{H}_{K0}+\hat{H}'_{d}+\hat{H}_{cp}\,,
\ee
where $H_{K0}$ describes the Hamiltonian of the Kondo spin plus site $\mu=0$, $\hat H'_d$ is the Hamiltonian of a ``reduced'' dot with $N-1$ sites $\mu=1,\ldots,N-1$  and $\hat{H}_{cp}$ contains the remaining coupling terms. Writing $\hat{\mathbf S}_{d}=\hat{\mathbf S}'_d + \hat{\mathbf s}_0$, where $\hat{\mathbf S}'_d = \sum_{\mu=1}^{N-1} \hat{{\bf s}}_\mu$  is the spin of the reduced dot, we have
\beqn\label{eq:2ca}
\hat{H}_{K0}&=&\varepsilon^c_0\,\hat{n}_0 -J_s\,\hat{\mathbf s}_0^2 + J_k\,\hat{\mathbf S}_K\cdot\hat{\mathbf s}_0\\
\hat{H}'_{d}&=&\hat{H}'_0-J_s\hat{\mathbf
  S}'^2_d\label{eq:2cb}\\\label{eq:2cc}
\hat{H}_{cp}&=&-\,2J_s\,\hat{\mathbf s}_0\cdot\hat{\mathbf S}'_d
+\sum_\sigma \left(t_{0}\,\hat{c}^\dagger_{0,\sigma}
\hat{c}^{\phantom{\dagger}}_{1,\sigma}+\,{\rm
  h.c.}\right).\quad\label{eq:2c3}
\eeqn
$\hat H'_0$ is the ``bare'' Hamiltonian of the reduced dot, given by
expressions similar to Eqs.~(\ref{eq:2a}) and (\ref{eq:2b}) but with
the sums over $\mu$ starting at $\mu=1$.

Here and in the following, operators in the reduced dot space of $N-1$ chain sites $\mu=1,\ldots,N-1$ are denoted by primed quantities. For such operators, the summation over sites $\mu$  starts from $\mu=1$ rather than from $\mu=0$.

\subsection{Site basis with good spin quantum numbers}

The Hamiltonian  $\hat{H}$ in Eq.~(\ref{eq:2bbb}) is invariant under
spin rotations and therefore conserves the total spin of the system
(Kondo spin plus dot spin) $\hat{\mathbf S}_{\rm tot} = \hat{\mathbf S}_K +\hat{\mathbf S}_d= \hat{\mathbf S}_{K0} +\hat{\mathbf S}'_d$. To take advantage of this symmetry, it is convenient to use a basis for which both the total spin $S_{\rm tot}$ and the corresponding magnetic quantum number $M_{\rm tot}\equiv S^z_{\rm tot}$ are good quantum numbers.

There are different ways to construct a basis with good total spin,
but one of them is particulary useful in the strong-coupling limit
$J_k \gg t_0, J_s$. To zeroth order in $t_0/J_k$ and $J_s/J_k$, we can ignore the coupling term $\hat H_{cp}$, in which case the subsystem of Kondo spin plus site 0 decouples from the reduced dot. The Hamiltonian $\hat H_{K0}$ is easily diagonalized by coupling the spins $\hat{\mathbf S}_{K}$ and $\hat{\mathbf s}_{0}$ to $\hat{\mathbf S}_{K0}\equiv\hat{\mathbf S}_{K}
+\hat{\mathbf s}_{0}$ and by using
$\hat{\mathbf S}_K \cdot \hat{\mathbf s}_0 = \left(\hat{\mathbf
    S}_{K0}^2 - \hat{\mathbf S}_K^2 - \hat{\mathbf s}_{0}^2 \right)/2$.

If site $\mu=0$ is singly occupied, i.e., $n_0=1$, this spin coupling will lead to either a singlet $S_{K0}=0$ (lowest energy) or a triplet $S_{K0}=1$ (highest energy).  However, if site $\mu=0$ is empty
$(n_0=0)$ or doubly occupied  $(n_0=2)$, the spin at site 0 and the
corresponding Kondo coupling term vanish. This results in an
unscreened Kondo spin in a doublet state ($S_{K0}=1/2$), the
energy of which is intermediate between the singlet and triplet states.

We now construct a basis of good total spin that also reflects the division into singlet/doulet/triplet manifolds. The eigenstates of $\hat H_{K0}$ are characterized by $S_{K0}, M_{K0}$ with $M_{K0}$ being the magnetic quantum number of $\hat{\mathbf S}_{K0}\,$. The eigenstates of the reduced dot Hamiltonian $\hath'_d$ with $N-n_0$ electrons are characterized by $|\gamma' S'_d M'_d \rangle$, where $S'_d,\,M'_d$ are the spin and spin projection, respectively, of the reduced dot and $\gamma'$ denotes all other quantum numbers distinguishing between states of the same $S'_d\,.$ We then couple the above eigenstates of $\hat H_{K0}$ with the eigenstates of the reduced dot to form states with good total spin and spin projection quantum numbers $S_{\rm tot},M_{\rm tot}\,$.
This basis of the coupled system is given by $|n_0, S_{K0}; \gamma',
S'_d; S_{\rm tot},M_{\rm tot}\rangle$. To keep the notation simple, we
omitted the quantum numbers $S_K=1/2$ and $s_0=n_0(2-n_0)/2$.

Spin selection rules determine the allowed values of the reduced dot
spin $S'_d$ for a given value of the total spin $S_{\rm tot}$. We have $S'_d=S_{\rm tot}$ for the singlet subspace, $S'_d=S_{\rm tot}\pm 1/2$ for the doublet subspace, and $S'_d=S_{\rm tot}, S_{\rm tot}\pm 1$ for the triplet subspace.

\section{Strong-Coupling Hamiltonian}

The strong-coupling limit is defined by $J_k \gg t_0$. Since $t_0 \sim
D \sim N \deltab$, this limit corresponds to $J_k \bar \rho \gg 1$,
where $\bar \rho=1/(N\deltab)$ is the average single-particle level
density per site. In the strong-coupling limit, the lowest eigenstates
of $\hat H$ are dominated by the singlet manifold. The bare singlet
Hamiltonian (in the limit when $\hat H_{cp}$ is ignored) is given by
the reduced dot Hamiltonian $\hat H_d'$ with $N-1$ electrons (except for a constant shift). However, virtual transitions between the singlet and doublet/triplet manifolds add correction terms to this Hamiltonian. Our goal is to determine the resulting effective Hamiltonian for the reduced dot in the strong-coupling limit.

\subsection{Projection technique}

In the limit of strong but finite Kondo coupling, the above three
manifolds (singlet, doublet and triplet) are coupled to each
other. Specifically, the exchange coupling term in $\hat H_{cp}$
couples the singlet and triplet manifolds, while the hopping term
between sites 0 and 1 couples each of the singlet and triplet
manifolds to the doublet manifold.  To account for these couplings we define projection operators $\hat{P}_s/\hat{P}_d/\hat{P}_t$ on the
corresponding singlet/doublet/triplet subspaces
($\hat{P}_s+\hat{P}_d+\hat{P}_t=1$)
and decompose the wave function $\psi=\psi_s+\psi_d+\psi_t$
accordingly.\cite{pietnotes}
The Schr\"odinger equation for the coupled system
$\hat{H}\psi=E\,\psi$ can then be written as
\be\label{eq:3}
\sum_{\beta=s,d,t}\,\hat{H}_{\alpha\beta}\,\psi_{\beta}=
E\,\psi_{\alpha}\,,
\ee
where each of the two indices $\alpha,\beta$ assumes any of three
values $\{s,d,t\}$ and
$\hat{H}_{\alpha\beta}\equiv\hat{P}_\alpha\,\hat{H}\,\hat{P}_\beta$.

In the strong-coupling limit, our system is described to zeroth
order by the singlet Hamiltonian $\hath_{ss}$, which contains
the bare reduced dot Hamiltonian  $\hath'_d$ (for
$N\!-\!1$ electrons) and $\hath_{K0}$ (which assumes a constant value), completely decoupled from each other. Higher order corrections come from the coupling terms in $\hat H_{cp}$ which lead to an effective ``dressed'' Hamiltonian of the reduced dot. This effective Hamiltonian $\hat H^{\rm eff}$ is formally determined by eliminating $\psi_d$ and $\psi_t$ in Eqs.~(\ref{eq:3}) and by writing a single equation in the singlet manifold $\hat{H}^{\rm eff} \psi_s = E\psi_s$, where
$\hat{H}^{\rm eff}$ is given by
\begin{multline}
\hat{H}^{\rm eff}= \hat{H}_{ss}+\hat{H}_{st}\left(E-\hat{H}_{tt}\right)^{-1}
\hat{H}_{ts}
\label{eq:4}\\
+\left[\hat{H}_{sd}+\hat{H}_{st}\left(E-\hat{H}_{tt}\right)^{-1}\hat{H}_{td}
\right]\\
\times\left\{E- \left[\hat{H}_{dd}+\hat{H}_{dt}\left(E-
\hat{H}_{tt}\right)^{-1}\hat{H}_{td}\right]\right\}^{-1}\\
\times \left[\hat{H}_{ds}+\hat{H}_{dt}\left(E-\hat{H}_{tt}
\right)^{-1}\hat{H}_{ts}
  \right] \,.
\end{multline}

The diagonal components $\hat{H}_{\alpha\alpha}$  in the above equation are determined by evaluating $\hat H _{K0}$ [Eq.~(\ref{eq:2ca})] in each of the three
subspaces $\alpha=\{s,d,t\}$. The coupling terms in $\hath_{\rm cp}$
[Eq.~(\ref{eq:2cc})] do
not contribute to these diagonal components with the exception of
$\hat{\mathbf s}_0\cdot\hat{\mathbf S}'_d$ which contributes to $\hat H_{tt}$ only. We find

\beqn\label{eq:5}
\hat{H}_{ss}\!&=&\!\hat{P}_s\Big(\vare^c_0
-\frac{3\,J_k}{4}-\frac{3\,J_s}{4}+\hat H_d' \Big)\hat{P}_s\,,\\
\hat{H}_{dd}\!&=&\!
\hat{P}_d\Big( \vare^c_0\,\hat{n}_0  + \hat H_d' \Big)\hat{P}_d\,,\\
\hat{H}_{tt}\!&=&\!\hat{P}_t\Big(\vare^c_0 +\frac{J_k}{4}-\frac{3\,J_s}{4}+\hat H_d' -2J_s\,\hat{\mathbf S}_0\cdot\hat{\mathbf S}'_d \Big)\hat{P}_t\,. \label{eq:5c}\quad\quad
\eeqn

Contributions to off-diagonal components $\hat{H}_{\alpha\beta}$ with $\alpha\neq\beta$ originate in $\hat H_{\rm cp}$. The hopping term in $\hat H_{\rm cp}$ changes the spin $S_{K0}$ by $\pm 1/2$ and can only couple the doublet to each of the singlet and triplet manifolds, while the exchange term $\hat{\mathbf s}_0\cdot\hat{\mathbf S}'_d$ in $\hat H_{\rm cp}$ only couples the singlet and triplet manifolds.

\subsection{Expansion in the strong-coupling limit}

The effective Hamiltonian $\hat{H}^{\rm eff}$ and the construction of a good spin basis in the previous subsection are exact, in
that no approximations were made beyond the original Hamiltonian $\hath$ in Eq.~(\ref{eq:1}). However, the form (\ref{eq:4}) of
$\hat{H}^{\rm eff}$ is not very useful for practical
calculations. In the strong-coupling limit $J_k \gg t_0 \sim N\deltab$. Since the exchange constant $J_s$ is typically below $\sim \deltab$, the condition $J_k \gg J_s$ is automatically satisfied in the strong-coupling limit.  We can then expand $\hat{H}^{\rm eff}$ in the two small dimensionless parameters $t_0/J_k \sim 1/(J_k \bar \rho)$ and $J_s/J_k$. We will do so up to fourth order in these parameters, where the expansion terms are measured in units of $J_k$ (this energy unit is set by the energy of the singlet).

The starting point for this expansion is the unperturbed singlet
Hamiltonian $\hath_{ss}$, the eigenbasis of which is given by
$|n_0=1, S_{K0}=0; \gamma', S'_d; S_{\rm tot}=S'_d, M_{\rm
  tot}\rangle$. The corresponding eigenvalues are
\be\label{eq:5b}
E^{(0)}_{m}=
-\frac{3}{4}(J_k+J_s) +\vare^c_0 + E'^{(0)}_m-J_s\,S_{\rm tot}(S_{\rm tot}+1)\,,
\ee
where $E'^{(0)}_m$ are the eigenvalues of $\hath'_0$.
These unperturbed eigenvalues, $E^{(0)}_{m}$, are the limiting
solutions to which the eigenvalues $E_m$ of the
full Hamiltonian in Eq.~(\ref{eq:4}) converge for $J_k\to\infty$.
The differences between $E^{(0)}_{m}$ and $E_m$ at large but finite values of $J_k$ are induced by the virtual transitions from the singlet to the doublet or triplet subspaces. These
virtual excitations, in turn, give rise to effective
interaction terms in the reduced dot, denoted by
 $\delta\hath^{\rm eff}$. The full effective Hamiltonian in the singlet manifod is then given by $\hath^{\rm eff}=\hath_{ss}+\delta\hath^{\rm  eff}$.

  The effective interaction terms in $\delta\hath^{\rm eff}$ must be
consistent with charge and spin conservation.\cite{hewson93} In particular, $\delta\hath^{\rm eff}$ must be a scalar operator in spin space (i.e., invariant under rotations in spin space) and invariant under time reversal. This restricts the possible terms that appear in the effective Hamiltonian in the strong-coupling limit.

Scalar one-body terms, i.e., $\hat n_1$ and $(\sum_\sigma\hat{c}^\dagger_{1,\sigma}
\hat{c}^{\phantom{\dagger}}_{2,\sigma}+\,{\rm h.c.})$ lead to a
renormalization of the one-body part of the reduced dot Hamiltonian
$\hath'_d$.  Two-body scalars that can be constructed from the spin
$\hat{\mathbf s}_1$ at site $1$ and the total spin of the reduced dot
$\hatsb'_d$ are $\hat{\mathbf s}^2_1$, $\hat{\mathbf
  s}_1\cdot\hatsb'_d$ and $\hatsb'^2_d$.  The first $\hat{\mathbf
  s}^2_1=\frac{3}{4}\hat n_1(2-\hat n_1)$ is the Nozi\`eres term known
from the conventional Kondo problem\cite{nozi} (in the absence of
exchange, $J_s=0$) but the other two terms are new. The scalar triple product $i\hat{\mathbf S}'_d\cdot(\hat{\mathbf s}_1\times\hat{\mathbf S}'_d)$ (the imaginary $i$ is necessary for time-reversal invariance) does not lead to a new term since $i\hat{\mathbf S}'_d\cdot(\hat{\mathbf s}_1\times\hat{\mathbf S}'_d)=
-\,\hat{\mathbf s}_1\cdot\hat{\mathbf S}'_d$, while a fourth order invariant is given by $\hatsb'^4_d$. Other invariants such as
$\hatsb'_d\cdot\,\sum_{\sigma,\sigma'}\! \left(\hat{c}^\dagger_{1,\sigma}\,
\boldsymbol{\tau}_{\sigma,\sigma'}\,
\hat{c}^{\phantom{\dagger}}_{2,\sigma'}
\right)+\,{\rm h.c.}$, $\hatsb'_d\cdot\,\sum_{\sigma,\sigma'}\!
\left(\hat{c}^\dagger_{2,\sigma}\,
\boldsymbol{\tau}_{\sigma,\sigma'}\,
\hat{c}^{\phantom{\dagger}}_{1,\sigma'}
\right)+\,{\rm h.c.}$ and $\hat {\mathbf s}_1\cdot\hatsb'_d\,\sum_\sigma\Big(\hat{c}^\dagger_{1,\sigma}
\hat{c}^{\phantom{\dagger}}_{2,\sigma}+\, \hat{c}^\dagger_{2,\sigma}
\hat{c}^{\phantom{\dagger}}_{1,\sigma} \Big)+\,{\rm h.c.}$ are  allowed but, as we shall see, they cancel out in the effective Hamiltonian.

We rewrite the effective Hamiltonian in (\ref{eq:4}) as $\hat H^{\rm eff} = \hat H_{ss} + \sum_{i=1}^4 \delta \hat H_i$ where
\begin{eqnarray}\label{H-eff}
\delta \hat H_1 &=& \hat{H}_{sd}\,\frac{1}{(E-\hat{H}_{dd})+\hat{H}_{dt}\,\frac{1}{E-
\hat{H}_{tt}}\,\hat{H}_{td}}\,\hat{H}_{ds} \;, \nonumber\\
\delta \hat H_2 &= & \hat{H}_{st}\,\frac{1}{E-\hat{H}_{tt}}\,\hat{H}_{ts} \;, \\
\delta \hat H_3 &\approx& \hath_{st}\,\frac{1}{E-\hath_{tt}}\,\hath_{td}\,\frac{1}{E-\hath_{dd}}\,
\hath_{ds}+\,{\rm h.c.} \;,\nonumber \\
\delta \hat H_4& \approx &\hath_{st}\frac{1}{E-\hath_{tt}}\hath_{td}\frac{1}{E-\hath_{dd}}
\hath_{dt}\frac{1}{E-\hath_{tt}}\hath_{ts}\,.\nonumber
\end{eqnarray}
In the terms $\delta \hat H_3$ and $\delta
 \hat H_4$ above we have replaced $\hat{H}_{dd}+\hat{H}_{dt}\left(E-
\hat{H}_{tt}\right)^{-1}\hat{H}_{td}$ by $\hat{H}_{dd}$ (the neglected term gives contributions that are higher than fourth order in the expansion parameters).

We next expand each $\delta \hat H_i$ to fourth order in the
parameters $t_0/J_k$ and $J_s/J_k$. Since the energy $E$ is of the
order $J_k$, the fractions appearing in each term can be brought to a form $1/(1-\hat{X})$ with $\hat{X}$ being small in the expansion
parameters. We then approximate $1/(1-\hat{X})\approx 1+\hat{X} + \hat X^2$. In the following we summarize the explicit calculation of each term.

\subsubsection{Evaluation of\, $\delta \hat H_1$}

For $\delta \hat H_1$ we find
\be\label{eq:h1}
\delta \hat H_1\approx -\frac{4}{3J_k}\hat{H}_{sd}\left( 1+ \hat A +\hat B + \hat C +\hat A^2 +\hat B^2\right)\hat{H}_{ds} \;,
\ee
where
\begin{eqnarray}\label{dh1}
\hat A & = & \frac{4}{3 J_k}\left[E'^{(0)}-\hath'_0 + \vare^c_0\,(1-\hat{n}_0) \right] \;, \\
\hat B & = & \frac{4}{3 J_k} J_s [-3/4-S_{\rm tot}(S_{\rm tot}+1)+\hat{\mathbf S}'^2_d] \;,\\
\hat C & = & \frac{4}{3 J^2_k} \hath_{dt}\,\hath_{td}\,.
\end{eqnarray}
In Eq.~(\ref{eq:h1}), we omitted the product terms $\hat A\hat C,\,\hat B
\hat C,\, \hat C^2$ since their contribution is higher than fourth
order, while the contribution of $\hat A \hat B + \hat B \hat A$ can
be shown to vanish identically.

To keep track of the various contributions for each of the $\delta
\hat H_i$, we label them in the following by $\delta \hat
H_{i,j}$. These terms are understood to act only in the space of the
reduced dot while the Kondo spin and the spin on site 0 are locked into a singlet. The operators $\delta \hat H_{i,j}$ in the reduced dot space are obtained by taking a partial expectation value $\langle \ldots \rangle_s$ in the singlet state. The corresponding operators in the full space  are given, respectively, by $P_s \langle \ldots\rangle_s P_s$. In Appendix A we list several relations that are useful in calculating the expectation values of various operators in the singlet space.

The most dominant contribution to Eq.~(\ref{eq:h1}) arises from the unity operator term (in the round brackets). We find
\begin{eqnarray}
 \delta \hat H_{1,1}  \equiv  -{4\over 3 J_k
 }\langle\hat{H}_{sd}\hat{H}_{ds} \rangle_s & = &  -{4\over 3 J_k }\,
 \langle \hath^{(0,1)}_{\rm hop} \hath^{(0,1)}_{\rm hop}\rangle_s  \nonumber \\ & = &
-{4\over 3 J_k }\,|t_{0}|^2 \;, \label{eq:h11}
\end{eqnarray}
where we have substituted $\hat H_{sd}$ by the hopping Hamiltonian between sites $0$ and $1$,
\be\label{hop-01}
\hath^{(0,1)}_{\rm hop} = \sum_\sigma t_{0}\,\hat{c}^\dagger_{0,\sigma}
\hat{c}^{\phantom{\dagger}}_{1,\sigma}+ \,{\rm h.c.}\,,
\ee
and used Eq.~(\ref{eq:a7'}).  Alternatively,
$\hat{H}_{sd}\hat{H}_{ds}$ describes the spin transitions illustrated in Fig.~\ref{fig:2}a, and the result in Eq.~(\ref{eq:h11}) can be
derived by using Table~\ref{tab:1} in Appendix A to sum up all the corresponding transition pathways.

\begin{figure}[!b]
\includegraphics[angle=0,width=85mm,clip=true]{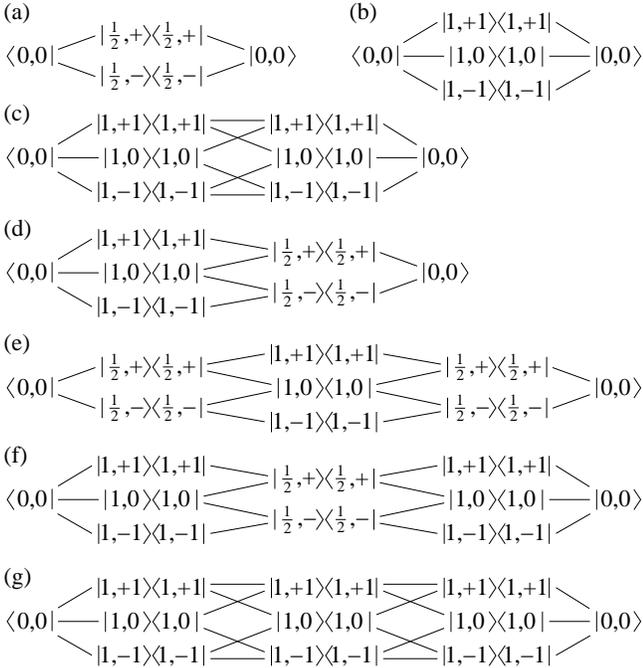}
\caption{Spin transition diagrams used
 to derive the effective strong coupling Hamiltonian in
Eq.~(\ref{eq:effective}). All transitions connect
two singlet states, characterized
by the quantum numbers $S_{K0}=0,\,S_{K0,z}=0$. The intermediate
transition pathways involve combinations of doublet
$\left(S_{K0}=1/2,\,S_{K0,z}=\pm 1/2\right)$
and triplet $\left(S_{K0}=1,\,S_{K0,z}=\pm 1,\,0\right)$
states.}\label{fig:2}
\end{figure}

The term containing $\Ah$ in Eq.~(\ref{eq:h1}) yields
corrections that are second order in $t_0/J_k$.
Using the difference in the values of $\hath'_0$, $\hat{n}_0$
in the singlet and doublet subspaces, and Eqs.~(\ref{eq:a6a})--(\ref{eq:a6b}), we find
\begin{multline}\label{eq:11}
\delta \hat H_{1,2}\equiv -\frac{4}{3J_k}\,
\langle \hat{H}_{sd}\Ah\hat{H}_{ds} \rangle_s\\
 = -\left(\frac{4}{3J_k}\right)^2|t_{0}|^2\,
\Bigg[(\varepsilon^c_0-\varepsilon^c_1)
+(\varepsilon^c_1-\varepsilon^c_0)\,\hat{n}_1\,+\\
+\frac{1}{2}\sum_\sigma \left(t^{\phantom{\dagger}}_{1}\,
\hat{c}^\dagger_{1\sigma}
\hat{c}^{\phantom{\dagger}}_{2\sigma}+
\,{\rm h.c.}
\right)\Bigg] \;.
\end{multline}
Except for an additional constant shift of $-\left(\frac{4}{3J_k}\right)^2|t_{0}|^2\,(\varepsilon^c_0-\varepsilon^c_1)$, this is a one-body operator that can be incorporated into the unperturbed singlet basis by simply redefining the site energy
$\varepsilon^c_1$ and hopping amplitude
$t_1$ in the unperturbed Hamiltonian $\hath'_0$ as follows\cite{pietnotes}
\beqn\label{eq:13}
\varepsilon^c_1&\rightarrow&\varepsilon^c_1
-\left(\frac{4\,|t_{0}|}{3J_k}\right)^2\left(\varepsilon^c_1-\varepsilon^c_0
\right) \;,\\
t_{1}&\rightarrow&t_{1}\left[1-\frac{1}{2}
\left(\frac{4\,|t_{0}|}{3J_k}\right)^2\right] \,.\label{eq:13d}
\eeqn

The term involving $\Bh$ in Eq.~(\ref{eq:h1}) contributes only for a finite exchange interaction ($J_s\neq 0$). We have
\be\label{eq:14}
\langle \hat{H}_{sd}\,\hat{\mathbf S}'^2_d\,\hat{H}_{ds}\rangle_s =\frac{|t_{0}|^2}{2}\,
\sum_{\sigma}\left(\hat{c}^\dagger_{1\sigma}
\hat{\mathbf S}'^2_d
\hat{c}^{\phantom{\dagger}}_{1\sigma}+\hat{c}^{\phantom{\dagger}}_{1\sigma}
\hat{\mathbf S}'^2_d\hat{c}^{\dagger}_{1\sigma}
\right)\,.
\ee
Using the identities (\ref{eq:a7})--(\ref{eq:a7p}),
Eq.~(\ref{eq:14}) can be simplified to
give Eq.~(\ref{eq:a7a}) in the Appendix.
Using $S'_d=S_{\rm tot}$ in the singlet subspace, we obtain
\be\label{eq:16}
\delta \hat H_{1,3}\equiv -\frac{4}{3 J_k}\,
\langle \hat{H}_{sd}\Bh\hat{H}_{ds}\rangle_s\\ = 2\,J_s\left(\frac{4\,|t_{0}|}{3J_k}\right)^2\,\hat{\mathbf s}_1
\cdot\hat{\mathbf S}'_d \,.
\ee
We note that $\delta \hat H_{1,3}$ is a spin invariant in the reduced dot space.

The term involving $\Ch$ in Eq.~(\ref{eq:h1}) is given by
\be\label{del_H_14}
\delta \hat H_{1,4}\equiv-\frac{16}{9 J_k^3 } \langle
\hath_{sd}\hath_{dt}\hath_{td}\hath_{ds} \rangle_s \;.
\ee
 This term appears in the conventional Kondo problem (where $J_s=0$) and is known as the Nozi\`eres term. Nozi\`eres found\cite{nozi} that this term yields an effective interaction in the singlet-space that repels opposite spins on site 1. This term is induced by virtual transitions
of the type singlet--doublet--triplet--doublet--singlet. Once we insert a triplet projection $\hat P_t$  in the r.h.s.~of Eq.~(\ref{del_H_14}),
i.e., we write the corresponding singlet expectation value as
$\langle\hath_{sd}\hath_{dt}\hat P_t\hath_{td}\hath_{ds} \rangle_s$,
we can replace both $\hat H_{sd}$ and $\hat H_{dt}$ by the hopping
Hamiltonian $\hath^{(0,1)}_{\rm hop}$ between sites $0$ and $1$ [see
Eq.~(\ref{hop-01})]. Using $\hath^{(0,1)}_{\rm hop} \hat P_t
\hath^{(0,1)}_{\rm hop} = \hath^{(0,1)}_{\rm hop} \hath^{(0,1)}_{\rm
  hop} - \hath^{(0,1)}_{\rm hop} \hat P_s \hath^{(0,1)}_{\rm hop}$, we have
\begin{eqnarray}
\langle\hath_{sd}\hath_{dt}\hath_{td}\hath_{ds} \rangle_s = &
\langle \hat H^{(0,1)}_{\rm hop}\hat H^{(0,1)}_{\rm hop} \hat
H^{(0,1)}_{\rm hop}\hat H^{(0,1)}_{\rm hop} \rangle_s \nonumber \\ & -
|\langle \hat H^{(0,1)}_{\rm hop} \hat H^{(0,1)}_{\rm hop} \rangle_s|^2 \;.
\end{eqnarray}
 With the help of Eqs.~(\ref{eq:a1'}), (\ref{eq:a2'}), (\ref{eq:a7'}) and (\ref{eq:a6a}), we then find
\be\label{noz}
\delta \hat H_{1,4}= -3\left(\frac{4}{3 J_k }\right)^3|t_{0}|^4\,\hat{\mathbf s}^2_1 = -\frac{16}{3 J_k^3 }\,|t_{0}|^4 \,\hat n_1(2-\hat n_1)\,\;.
\ee
An alternative way to calculate $\delta \hat H_{1,4}$ is to use the spin diagram in Fig.~\ref{fig:2}e. It can be reduced to the transition diagram in Fig.~\ref{fig:2}b with the help of Table \ref{tab:1} in Appendix A.

The Nozi\`eres term (\ref{noz}) vanishes when site $1$ is either empty ($n_1=0$) or doubly occupied ($n_1=2$) but is negative for $n_1=1$, thus favoring a singly occupied site $1$.

The contribution from $\hat A^2$ in Eq.~(\ref{eq:h1}) is evaluated using Eq.~(\ref{eq:a6c}) and  Eq.~(\ref{eq:a6d}) and leads to a constant shift
\begin{multline}
 \delta \hat H_{1,5}\equiv -\frac{4}{3J_k}\,
\langle\hat{H}_{sd} \hat A^2\hat{H}_{ds}\rangle_s\\ = -\left({4\over 3 J_k}\right)^3 |t_{0}|^2\big[|t_{1}|^2+\varepsilon^2_0
+\varepsilon^2_1-2\varepsilon^{\phanda}_0
\varepsilon^{\phanda}_1\big] \;.
\end{multline}
Finally, the contribution from $\hat B^2$ in Eq.~(\ref{eq:h1}) is found to be
\be
\delta \hat H_{1,6}\equiv -\left({4\over 3 J_k}\right)^3|t_{0}|^2J^2_s
\big(\hat{\mathbf S}'^2_d+2\,\hat{\mathbf s}_1\cdot
\hat{\mathbf S}'_d\big)\;,
\ee
 where we have used Eq.~(\ref{eq:a7ap}).

\subsubsection{Evaluation of\, $\delta \hat H_2$}

We next turn to the singlet-triplet transitions as described by
$\delta \hat H_2$ in Eq.~(\ref{H-eff}). The corresponding expression for $\delta \hat H_2$ is given by
\begin{equation}\label{eq:17a}
\delta \hat H_2\approx-\frac{1}{J_k}
\hat{H}_{st}\left(1+\hat D+\hat E+\hat E^2\right)\hat{H}_{ts}\,,
\end{equation}
where
\begin{eqnarray}\label{eq:17b}
\hat D & = & \frac{1}{J_k}\left[E'^{(0)}-\hath'_{0}\right]\,,\\
\hat E & = & \frac{J_s}{J_k}\left[-S_{\rm tot}(S_{\rm tot}+1)
+\hat{\mathbf S}'^2_d+2\,\hat{\mathbf s}_0\cdot\hat{\mathbf
  S}'_d\right]\,.
\end{eqnarray}
In Eq.~(\ref{eq:17a}) we omitted the terms $\hat D^2$ and $\hat D \hat E$, which can be shown to vanish.

The dominating term in  Eq.~(\ref{eq:17a}) is the one involving the
unity operator. The corresponding term induces a spin transition as in
Fig.~\ref{fig:2}b. Using $\hath_{st}=-2J_s\hatp_s(\hat{\mathbf
  s}_0\cdot\hat{\mathbf S}'_d)\hatp_t$  and Eq.~(\ref{eq:a9}) we find
\be\label{eq:18}
\delta\hat H_{2,1}\equiv-\frac{1}{J_k}
\langle\hat{H}_{st}\hat{H}_{ts}\rangle_s= -\frac{J^2_s}{J_k}\,\hatsb'^2_d\,.
\ee
The same result can also be obtained with the help of Table \ref{tab:2} in Appendix A. The contribution $\delta \hat H_{2,2}$ induced by the term involving $\Dh$ in Eq.~({\ref{eq:17a}}) can be simplified by using
$\langle\hath_{st}\hath'_{0}\hath_{ts}\rangle_s=J^2_s\,\hatsb'_d
\hath'_{0}\hatsb'_d$. Since  $\hatsb'_d$ commutes with the scalar operator $\hath'_{0}$, we find
\be
\delta\hat H_{2,2}\equiv-\frac{1}{J_k}
\langle\hat{H}_{st}\hat D\hat{H}_{ts}\rangle_s= 0\,.
\ee
The contribution $\delta\hat H_{2,3}$ induced by the
term $\Eh$ in Eq.~({\ref{eq:17a}})
can also be simplified since $\hatsb'^2_d$ acts in the reduced dot space (and therefore has identical action in the singlet and triplet manifolds). The resulting  expression gives rise to transition pathways as shown in Fig.~\ref{fig:2}c, the sum over which is further simplified using (\ref{eq:a10'}) to give
\be\label{eq:19}
\delta\hat H_{2,3}\equiv -\frac{1}{J_k}
\langle\hat{H}_{st}\hat E\hat{H}_{ts}\rangle_s=
\frac{J^3_s}{J^2_k}\,\hatsb'^2_d\,.
\ee

The last term $\delta H_{2,4}$ in Eq.~(\ref{eq:17a}), containing $\Eh^2$, is found to be $\delta H_{2,4}=-(16J^4_s/J^3_k)\,
\big(\hat{\mathbf s}_0\cdot\hat{\mathbf S}'_d\big)^4$ and corresponds to the transition diagram in Fig.~\ref{fig:2}g.
Since $\hat H_{st}$, as a scalar operator, commutes with $\hat{\mathbf S}'^2_d$, all other terms vanish identically.
Using Eqs.~(\ref{eq:a10a}) this expression can be simplified to give
\be\label{eq:20}
\delta\hat H_{2,4}\equiv-\frac{1}{J_k}
\langle\hat{H}_{st}\,\hat E^2\,\hat{H}_{ts}\rangle_s=
-\frac{J^4_s}{J^3_k}\,\hat{\mathbf S}'^2_d\,.
\ee

\subsubsection{Evaluation of\, $\delta \hat H_3$}

Following Eq.~(\ref{H-eff}), the subsequent contribution,
$\delta \hat H_3$, involves transitions to both the doublet and the triplet subspaces
\be\label{eq:21}
 \delta \hat H_3\approx
\frac{4}{3J^2_k}
\hat{H}_{st}\!\left(1+\hat D+\hat E\right)\!
\hath_{td}\!\left(1+\hat A + \hat B \right)\!\hath_{ds}+{\rm h.c.},
\ee
where the operators $\hat A-\hat E$ are the same as introduced above. Contributions involving products between $\{\hat D,\hat E\}$ and
$\{\hat A, \hat B\}$, respectively, are higher than fourth order in the expansion parameters and therefore not considered here.

The dominant contribution, $\delta\hat H_{3,1}$ originates in the transitions shown in Fig.~\ref{fig:2}d. Using Table \ref{tab:1} we can, however, simplify this transition diagram to the one of Fig.~\ref{fig:2}b, for which we obtain
\be\label{eq:22}
\delta\hat
H_{3,1}\equiv\frac{4}{3J^2_k}\langle\hath_{st}\hath_{td}\hath_{ds}\rangle_s +\,{\rm h.c.}=
\frac{16J_s}{3J_k^2}|t_{0}|^2\,\hat{\mathbf s}_1\cdot\hatsb'_d\,.
\ee

To simplify the term $\delta\hat H_{3,2}$, involving $\Dh$ in
Eq.~(\ref{eq:21}), we make
use of $[\hath_{st},\hath'_{0}]=0$, leading to
\begin{eqnarray}
\delta \hat H_{3,2}&\equiv&\frac{4}{3J^2_k}\,\hath_{st}\,\hat D\,
\hath_{td}\,\hath_{ds}+\,{\rm h.c.}\label{eq:23}\\\nonumber
&=&\frac{8J_s}{3J^3_k}\,|t_{0}|^2\,
\left(E'^{(0)}-\hath'_{0}\right)\,\hat{\mathbf s}_1 \cdot \hatsb'_d +\,{\rm h.c.}
\end{eqnarray}
The diagonal matrix elements of $\delta \hat H_{3,2}$ (when evaluated in the eigenstates of the reduced dot $H_0'$) vanish. Off-diagonal matrix elements enter in second order perturbation theory (for the effective Hamiltonian) and would lead to a correction $\sim J^2_s|t_{0}|^4/J^6_k$ that is beyond the fourth order approximation. A convenient choice for $E'^{(0)}$ is the average energy of the two reduced dot eigenstates (appearing in the corresponding matrix element). This choice leads to
\be\label{eq:24}
\delta\hat H_{3,2}
= 0\,.
\ee

The next term $\delta\hat H_{3,3}$, produced by $\Eh$ in
Eq.~(\ref{eq:21}), can be brought into the form of
Fig.~\ref{fig:2}c (using Table \ref{tab:1}), where a sum of all occurring terms can be identified with Eq.~(\ref{eq:a10}),
yielding
\begin{eqnarray}
\delta\hat H_{3,3}&\equiv&\frac{4}{3J^2_k}\,\langle\hath_{st}\,\hat E\,
\hath_{td}\,\hath_{ds}\rangle_s+\,{\rm h.c.}\nonumber\\\label{eq:25}
&=& -\frac{8J^2_s}{3J^3_k}\,|t_{0}|^2\,\hat{\mathbf s}_1\cdot\hatsb'_d +\,{\rm h.c.}
\end{eqnarray}

The contribution $\delta\hat H_{3,4}$ induced by $\Ah$ in Eq.~(\ref{eq:21}), is calculated to be
\begin{eqnarray}\label{eq:26}
\delta\hat H_{3,4}&\equiv&\frac{4}{3J^2_k}\,\langle\hath_{st}\,
\hath_{td}\,\hat A\,\hath_{ds}\rangle_s+\,{\rm h.c.}\\\nonumber
&=&\frac{32J_s}{9J^3_k)}\,|t_{0}|^2\,
\hat{\mathbf s}_1\cdot\hatsb'_d\,\left(E'^{(0)}-\hath'_{0}\right)+\,{\rm h.c.}
\end{eqnarray}
Following the same arguments used for the evaluation of $\delta\hat
H_{3,2}$, we find $\delta\hat H_{3,4}=0$.

The term $\Bh$ in Eq.~(\ref{eq:21}) gives rise to
a transition diagram as in Fig.~\ref{fig:2}d which can be reduced to
a diagram as in Fig.~\ref{fig:2}b with the help of Table \ref{tab:3}.
Further simplifications involving several of the equations in Appendix A yield
\begin{eqnarray}
\delta\hat H_{3,5}&\equiv&\frac{4}{3J^2_k}\,\langle\hath_{st}\,
\hath_{td}\,\hat B\,\hath_{ds}\rangle_s+\,{\rm h.c.}\\
&=&\nonumber-\frac{J^2_s}{J_k}\left(\frac{4}{3J_k}\right)^2\,|t_{0}|^2\,
(\hatsb'^2_d+2\,\hat{\mathbf s}_1\cdot\hatsb'_d)+\,{\rm h.c.}
\end{eqnarray}

\subsubsection{Evaluation of\, $\delta \hat H_4$}
To determine the terms contributing to $\delta\hath_4$ (up to fourth order),
we can make the following approximations in Eq.~(\ref{H-eff}):
 $(E\!-\!\hath_{tt})^{-1}\approx -1/J_k$ and
$(E\!-\!\hath_{dd})^{-1}\approx -4/(3J_k)$. The resulting expression
for $\delta\hath_4$ corresponds to a transition diagram as in Fig.~\ref{fig:2}f. Using Eqs.~(\ref{eq:a1'}), (\ref{eq:a10}),  (\ref{eq:a9}) and (\ref{eq:a10'}) we find
\begin{eqnarray}
\delta\hat H_{4}&\approx&-\frac{4}{3J^3_k}\langle\hath_{st}\,
\hath_{td}\,\hath_{dt}\,\hath_{ts}\rangle\\
\nonumber&=&-\frac{4J^2_s}{3J^3_k}\,|t_{0}|^2\,
\left(\hatsb'^2_d+2\,\hat{\mathbf s}_1\cdot\hatsb'_d\right)\,.
\end{eqnarray}
Alternatively, we can also obtain this results by using Table I to
get a transition diagram as in Fig.~\ref{fig:2}c, which can
be simplified by using Eq.~(\ref{eq:a10}).

\subsubsection{Additional terms}

In the above calculations, we have replaced the energy eigenvalue
$E$ in Eq.~(\ref{eq:4}) by its unperturbed values $E^{(0)}$ as given by Eq.~(\ref{eq:5b}). However, additional terms to the effective Hamiltonian are found when corrections to $E^{(0)}$ are included self-consistently.  To the order we are interested in, it is sufficient to consider $\delta\hath_{1,1}$ and $\delta\hath_{2,1}$ [see Eqs.~(\ref{eq:h11}) and Eq.~(\ref{eq:18}), respectively] as corrections to  $E^{(0)}$
\be\label{eq:improvedapprox}
E\approx
E^{(0)} -\frac{4|t_{0}|^2}{3J_k} -\frac{J^2_s}{J_k}\,S_{\rm tot}(S_{\rm tot}+1)  \,.
\ee
Higher order corrections to $E$ in Eq.~(\ref{eq:improvedapprox})
lead to terms in the effective Hamiltonian that are higher than fourth order.

Adding now the two correction terms from Eq.~(\ref{eq:improvedapprox})
 to $E'^{(0)}$ in Eq.~(\ref{dh1}) gives the following term
\begin{eqnarray}
\delta\hath_{1,7} \equiv \left(\frac{4}{3J_k}\right)^3\,|t_{0}|^4\,
+\left(\frac{4J_s}{3J_k}\right)^2\frac{|t_{0}|^2}{J_k}\,\hatsb'^2_d\,.
\end{eqnarray}

A similar correction to $E'^{(0)}$ in Eq.~(\ref{eq:17b}) yields:
\begin{eqnarray}
\delta\hath_{2,5} \equiv \frac{4J^2_s}{3J^3_k}\,|t_{0}|^2\,\hatsb'^2_d+\frac{J^4_s}{J^3_k}
\,\hatsb'^4_d\,.
\end{eqnarray}
Up to fourth order, corrections to $E'^{(0)}$ do not lead to additional terms in  $\delta \hat H_3$ and $\delta \hat H_4$.

\subsection{Effective Hamiltonian: a complete expression}\label{sec:g}
Collecting all the terms found in the previous section, the effective Hamiltonian $\hath^{\rm eff}=\hath_{ss}+\sum^{4}_{i=1}\delta\hath_i$ is given (to fourth order in $t_0/J_k$ and $J_s/J_k$) by
\begin{multline}
\hath^{\rm eff}\approx
\eta +{\tilde{H}}_d' -\left(\frac{4}{3 J_k}\right)^3|t_{0}|^4\left(3\,\hat{\mathbf s}^2_1-1\right) \\
+J_k\left(\frac{80}{9}\frac{J_s}{J_k}-\frac{536}{27}
\frac{J^2_s}{J^2_k}
\right)\frac{|t_{0}|^2}{J^2_k}\,\hat{\mathbf s}_1\cdot
\hat{\mathbf S}'_d +\,\frac{J^4_s}{J^3_k}\,\hat{\mathbf S}'^4_d \;,
\label{eq:effective}
\end{multline}
where $\eta$ is a constant
\begin{eqnarray}
\eta & = & \vare^c_0-\frac{3\,(J_k+J_s)}{4}-\frac{4|t_{0}|^2}{3J_k}\\ & - & \left(\frac{4}{3J_k}\right)^3
|t_{0}|^2\,\Big[|t_{1}|^2+(\varepsilon^c_0)^2+(\varepsilon^c_1)^2
-2\,\varepsilon^c_0\,\varepsilon^c_1\Big]\;,\nonumber
\end{eqnarray}
and ${\tilde{H}}_d'$ describes a renormalized universal Hamiltonian for the reduced dot
\begin{equation}
{\tilde{H}}_d'= {\tilde{H}}_0' - \tilde{J}_s\hat{\mathbf S}'^2_d \;.
\end{equation}
${\tilde{H}}_0'$ is a renormalized one-body Hamiltonian of the
reduced dot, obtained from the original one-body Hamiltonian of the
reduced dot, $\hath_0'$, by redefining $\vare^c_1$ and $t_1$ according to Eqs.~(\ref{eq:13}) and (\ref{eq:13d}), respectively. This tridiagonal Hamiltonian can be rediagonalized
${\tilde{H}}_0'=\sum_{n=1, \sigma}^{N-1} \tilde{\varepsilon}^o_{\tilde{n}} \hat{a}^\dagger_{\tilde{n}\sigma} \hat{a}_{\tilde{n}\sigma}$ to define new effective single-particle orbitals $\hat{a}^\dagger_{\tilde{n}\sigma} | 0\rangle$ and energies  $\tilde{\varepsilon}^o_{\tilde{n}}$ of the reduced dot.
$\tilde{J}_s$ is a renormalized exchange constant
\begin{equation}\label{renorm-ex}
\tilde{J}_s= J_s \left(1+\frac{J_s}{J_k}-\frac{J^2_s}{J^2_k}+\frac{J^3_s}{J^3_k}
+\frac{112}{27}\,\frac{J_s\,|t_{0}|^2}{J^3_k}\right) \;.
\end{equation}
The most dominant contributions in (\ref{renorm-ex}) are positive and thus lead to a stronger exchange interaction in the reduced dot than in the original dot, $\tilde J_s>J_s$.  Since the Kondo spin and the spin at site $0$ are coupled to a singlet, the spin of the reduced dot $\hatsb'_d=\hatsb_{\rm tot}$, and is thus conserved (i.e.,
$S'_d=S_{\rm tot}$ and $M'_d\equiv S'_{d,z}=M_{\rm tot}$ are good quantum numbers).

The effective Hamiltonian of the reduced dot contains several
additional interaction terms, see Eq.~(\ref{eq:effective}). The term
proportional to
$(3\,\hat{\mathbf s}^2_1- 1)$ is the Nozi\`eres term, known from the
conventional Kondo problem (in the absence of exchange, $J_s=0$).\cite{nozi}  The term proportional to $\hat{\mathbf s}_1\cdot
\hat{\mathbf S}'_d $ is a new effective interaction in the reduced dot that is induced by the finite exchange interaction ($J_s\neq 0$) and describes an exchange interaction between the spin at site 1 and the spin of the reduced dot. This exchange interaction is to leading order antiferromagnetic but depending on the particular values of $J_k$ and $J_s$ it can also become ferromagnetic. The last term in (\ref{eq:effective}) is a four-body term but using $\hatsb'^4_d=\hatsb^4_{\rm tot}$, it is easily evaluated in terms of the conserved total spin $S_{\rm tot}$. It can be combined with the renormalized exchange interaction in the reduced dot by defining an  exchange coupling that depends on the total spin.

\section{Eigenvalues of the effective Hamiltonian}

The effective Hamiltonian of the reduced dot is valid to fourth order in $t_0/J_k$ and $J_s/J_k$ (when its terms are measured in units of $J_k$).  To determine its eigenenergies to this fourth order, it is sufficient to solve $\hath^{\rm eff}$ in {\em first-order} perturbation theory. Both $\tilde H_d'$ and $\hat{\mathbf S}'^4_d$ are diagonal in a basis of good orbital occupations and good total spin of the reduced dot, while a second-order perturbation theory of the remaining interaction terms leads to terms that are higher than fourth order in the combined power of $t_0/J_k$ and $J_s/J_k$.

As required by first-order perturbation theory, we
evaluate the expectation value of  $\hath^{\rm eff}$
in the unperturbed basis, i.e., in the eigenbasis of the
renormalized universal Hamiltonian $\tilde H_d'$ of the reduced dot.
For simplicity, we will denote the good spin eigenstates of $\tilde
H_d'$ by $|\xi\rangle$ and the corresponding expectation values by $\langle\ldots\rangle_\xi$. Most terms contained in $\hath^{\rm eff}$ [see Eq.~(\ref{eq:effective})] are diagonal in this basis, leaving only a few terms which require a special treatment.

Consider first the evaluation of the Nozi\`eres term. The calculation of $\langle \hat{\mathbf s}^2_1 \rangle_\xi$ simplifies for the lowest eigenstate of $\tilde H_d'$ at each given spin value $S'_d=S_{\rm tot}$. Those eigenstates of $\tilde H_d'$ with $M'_d=S'_d$ have a maximal spin projection
with only spin up electrons in singly occupied levels and thus have $\hat{n}^{\phantom{\dagger}}_{\tilde{n}\sigma}$ as good quantum numbers (in contrast to a general eigenstate of $\tilde{H}'_d$, where only the orbital occupation numbers $\hat{n}^{\phantom{\dagger}}_{\tilde{n}}$ are well defined). For these states we have
\begin{eqnarray}\label{eq:nozpert}
\langle\hat{\mathbf s}^2_1\rangle_\xi & = & \frac{3}{4}\,\sum_{\sigma=\pm}
\langle\hat{n}_{1\sigma}(1-\hat{n}_{1-\sigma})\rangle_\xi\\ \nonumber & =  & \frac{3}{4}\,\sum_{\sigma=\pm}
\langle\hat{n}_{1\sigma} \rangle_\xi
(1-\langle\hat{n}_{1-\sigma}\rangle_\xi)\,.
\end{eqnarray}

\begin{figure}[!t]
\includegraphics[angle=0,width=85mm,clip=true]{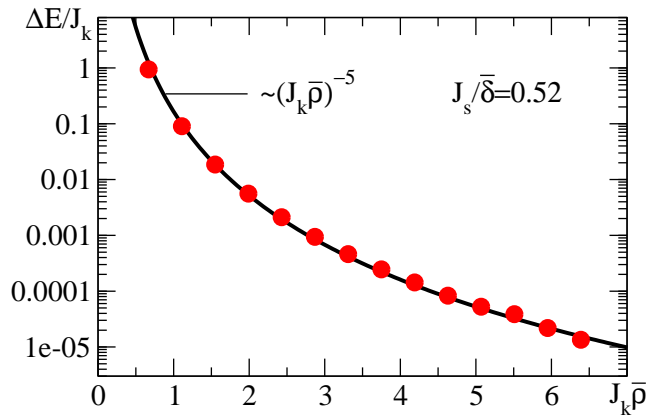}
\caption{(Color online) Energy difference $\Delta E$ between the
 estimate based on the effective Hamiltonian (\ref{eq:effective})
and the exact numerical result for the ground-state
energy in the subspace $S_{\rm tot}=3$ (energy is measured in units of $J_k$). The quantum dot contains 11 electrons in 11 spin-degenerate single-particle energy levels $\varepsilon^0_n$ with an arbitrary random matrix realization of the single-particle spectrum, $\varepsilon^0_{n}$, but with non-fluctuating orbital wave functions $\phi_n(0)=1/\sqrt{N}$.
The results, shown for an arbitrary but fixed value $J_s/\deltab =0.52$ (red symbols), behave like
$\sim 1/(J_k \bar \rho)^5$ (black solid line), expected for a strong-coupling expansion up to fourth order in $1/(J_k \bar\rho)$ in the limit $J_k\rhob\to\infty$.
}\label{fig:3}
\end{figure}

The occupation $\langle\hat{n}_{1\sigma}\rangle_\xi$ can be calculated from $\langle\hat{n}_{1\sigma}\rangle_\xi = \sum^{N-1}_{\tilde{n}=1}
|U'_{1\tilde{n}}|^2\,\langle \hat{n}_{\tilde{n}\sigma} \rangle_\xi$,
where $U'$ is the unitary matrix (of order $N\!-\!1$) transforming
between the renormalized single-particle orbitals of the reduced dot (with creation operators $\hat{a}^\dagger_{\tilde{n}\sigma}$)
and the site basis states $\mu= 1,\ldots,N-1$
\be\label{eq:uprimetrans}
\hat{c}^\dagger_{\mu,\sigma}=\sum_{\tilde{n}=1}^{N-1}U'_{\mu,\tilde{n}}\,
\hat{a}^\dagger_{\tilde{n},\sigma}\,.
\ee

For the good spin eigenstates of $\tilde{H}'_d$ with $S'_d=M'_d$, the
expectation value of $\hat{\mathbf s}_1\cdot\hat{\mathbf S}'_d$ is given by
\begin{eqnarray}\label{eq:27}
\langle\hat{\mathbf s}_1\cdot\hat{\mathbf S}'_d \rangle_\xi & = &(1+S'_d)\,\langle\hat{s}_{1,z} \rangle_\xi \\ \nonumber
& = &\frac{1+S'_d}{2}\,\sum_{\sigma=\pm}\sigma\,
\langle\hat{c}^\dagger_{1\sigma}
\hat{c}^{\phantom{\dagger}}_{1\sigma}\rangle_\xi \\ \nonumber
& = &\frac{1+S'_d}{2}\,\sum_{\sigma=\pm}\sum^{N-1}_{\tilde{n}=1}
\sigma\,|U_{1\tilde{n}}|^2\,\langle \hat{n}_{\tilde{n}\sigma}\rangle_\xi\,.
\end{eqnarray}

Using Eqs.~(\ref{eq:nozpert}) and (\ref{eq:27}), we can calculate the lowest many-body eigenenergy for each total spin value of the Kondo Hamiltonian (\ref{eq:1}) up to fourth order in  $t_0/J_k$ and $J_s/J_k$.

\section{Comparison with exact numerical diagonalization}

To validate our expression (\ref{eq:effective}) for the effective Hamiltonian, we compare our analytical results for the many-body energies in the strong-coupling limit with an
exact numerical diagonalization of the Hamiltonian (\ref{eq:1}) in a
good spin basis scheme we developed previously.\cite{tureci06,rotter08}

As a first test, we compare results for the lowest
energy of a given total spin (e.g., $S_{\rm tot}=3$). In Fig.~\ref{fig:3}, we show the difference $\Delta E$ (in units of $J_k$) between the energy determined from the effective Hamiltonian (\ref{eq:effective}) and the energy found from exact numerical diagonalization of (\ref{eq:1}) as a function of $J_k \bar \rho$ (at an arbitrary, but fixed value $J_s/\deltab=0.52$). This energy difference $\Delta E$ has to scale as $\propto 1/(J_k \bar \rho)^5$, which is the next order of correction in $1/(J_k \bar \rho)^n$ beyond the threshold of accuracy considered here. The results shown in Fig.~\ref{fig:3} confirm this scaling behavior and thereby the accuracy and completeness of our effective Hamiltonian. Similar results (not shown here) are found for other values of $S_{\rm tot}$ and for both even and odd number of electrons in the dot.

\begin{figure}[!t]
\includegraphics[angle=0,width=85mm,clip=true]{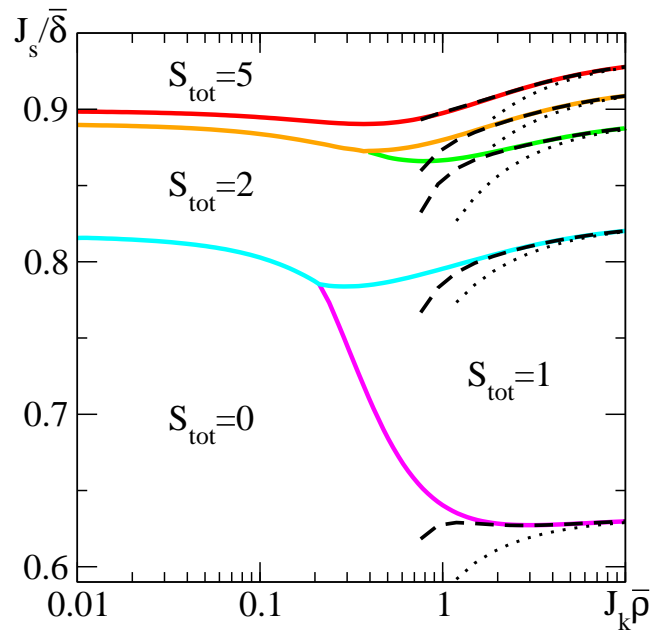}
\caption{(Color online) Ground-state spin $S_{\rm tot}$ of the
system in Fig.~\ref{fig:1} with finite exchange constant $J_s$ and Kondo coupling $J_k$. We consider 11 electrons in a dot with
$N=11$ single-particle levels, featuring an arbitrary random-matrix realization of the single-particle spectrum $\varepsilon^0_{n}$, but
non-fluctuating orbital wave functions $\phi_n(0)=1/\sqrt{N}$.
Lines show the transition curves separating regions of
fixed $S_{\rm tot}$. The estimates based on our strong-coupling
expansion [Eq.~(\ref{eq:effective})] for the Hamiltonian (dashed black lines) are compared with full numerical solutions (colored solid lines). The dotted black lines are the transition curves obtained when the strong-coupling limit is described by a renormalized universal Hamiltonian but without additional interaction terms [i.e., $\hath^{\rm eff}=\eta +\tilde H_d'$ in Eq.~(\ref{eq:effective})].
}
\label{fig:4}
\end{figure}

It is interesting to study the ground-state value of
the total spin $S_{\rm tot}$. This quantity was studied  theoretically\cite{murthy05,kaul06}
and can be probed experimentally.\cite{parallel} The ground-state spin $S_{\rm tot}$ undergoes successive transitions to higher values (known as the Stoner staircase) when the exchange coupling constant $J_s$ is varied between $J_s=0$ and a value $J_s\sim\deltab$ where the dot becomes fully polarized.  The transition steps in the Stoner staircase are shifted by the Kondo interaction. In Fig.~\ref{fig:4} we show numerical results for the ground state spin diagram in the two-dimensional parameter space of $J_s/\deltab,\,J_k \bar \rho$ for a particular mesoscopic realization of the single-particle Hamiltonian of the dot. The exact spin transition curves (colored lines) that separate regions of fixed ground-state spin $S_{\rm tot}$ are monotonically decreasing for $J_k\rhob\lesssim 1$ and monotonically increasing for $J_k\rhob\gtrsim 1$. Note also that for the particular mesoscopic realization chosen in Fig.~\ref{fig:4}, some values of $S_{\rm tot}$ (e.g., $S_{\rm tot}=1,\,3$)
never become the ground-state values of the total spin in the weak-coupling limit. In contrast, the ground-state spin assumes these values in the strong-coupling limit. Our analytical results in this limit, shown by the dashed lines in Fig.~\ref{fig:4}, are in very good agreement with the exact numerical results
down to values of $J_k\rhob\approx 2$. The dotted black lines in
Fig.~\ref{fig:4} are the corresponding transition lines when we do
not include interaction terms beyond the renormalized universal
Hamiltonian of the reduced dot, i.e., when we assume the effective
Hamiltonian to be just given as $\eta +\tilde H_d'$ [see Eq.~(\ref{eq:effective})]. These dotted curves converge much slower to the full numerical solutions (colored lines) than the dashed curves determined from the effective Hamiltonian (\ref{eq:effective}). However, both the dashed and the dotted lines reproduce the monotonic increase of the exact transition curves with $J_k$ for $J_k\rhob\gtrsim 2$. We conclude that this increase originates in the renormalization of the effective exchange coupling constant in the reduced dot, $\tilde{J}_s\approx J_s(1+J_s/J_k+\ldots)$, which is contained in the approximations used for both dashed and the dotted lines. The renormalized exchange constant $\tilde{J}_s$ decreases with increasing $J_k$, which implies, in turn, that the spin transition curves move upward with increasing $J_k$.

\section{Discussion and conclusion}

We have investigated the strong-coupling limit of the
Kondo problem when the screening electrons reside in a large quantum
dot that is described by the universal Hamiltonian. The novel feature
of this model, as compared with the conventional Kondo problem, is the
inclusion of discrete level spacings and electron-electron
interactions in the form of a ferromagnetic exchange interaction (that
is part of the universal Hamiltonian).

We have followed here a procedure that was originally
proposed in Ref.~\onlinecite{nozi} for the conventional Kondo problem in the absence of exchange correlations in the electron gas. As pointed out there, one can find the
effective Hamiltonian at strong Kondo coupling ($T \ll T_K$) by considering the \emph{bare} strong-coupling limit $J_k \gg t_0$. This bare strong-coupling limit is the one for which we have now provided a closed expression of all interaction terms up to fourth order in $t_0/J_k$ and $J_s/J_k$ when the electron gas is
described by the universal Hamiltonian. However, if the band width $D$ of this electron gas is very large $D\gg\bar{\delta}$, the limit of strong coupling can be effectively reached at much smaller values of $J_k$ than those of the bare limit. For such a system with large bandwidth, the strong-coupling limit corresponds to
a Kondo temperature $T_K$ that is larger than the system's temperature and average level spacing, $T_K\gg T,\bar{\delta}$. An important insight in Kondo theory is that the effective
Hamiltonians of both strong-coupling limits are related by a scaling analysis\cite{anderson70} in which the Kondo
Hamiltonian is renormalized by successive truncations of the band
width $D$, leaving the low-energy physics unchanged. As the reduced
band width $\tilde{D} \to 0$ (or equivalently $T \to 0$), the renormalized Kondo coupling constant $\tilde{J}_k$ diverges.\cite{wilson75} The coupling constants of the various terms in the effective strong-coupling Hamiltonian are typically determined by fitting the lowest excitations of the effective Hamiltonian with those obtained by a numerical solution of the full problem. For a detailed review of this procedure see, e.g., Refs.~\onlinecite{hewson93,nozi2,wilson75}.

By deriving in Eq.~(\ref{eq:effective}) the effective Hamiltonian  $\hath^{\rm eff}$ for the \emph{bare} strong-coupling limit,
we have completed successfully the first step in our goal to understand the strong-coupling limit of the Kondo problem in the presence of exchange correlations in the mesoscopic electron gas. We found that the exchange interaction in the universal Hamiltonian gives rise to two new terms in $\hath^{\rm eff}$: a four-body
contribution $\hat{\mathbf S}'^4_d$ that can be absorbed into a spin-dependent exchange coupling in the reduced dot, and a new
interaction term $\hat{\mathbf s}_1\cdot\hat{\mathbf S}'_d$
that describes an exchange interaction between an electron in the vicinity of the Kondo spin and the total spin of the reduced dot. This term is induced by the virtual polarization of the Kondo singlet involving excursions to both the doublet and triplet subspaces. Unlike the conventional Kondo problem ($J_s=0$), this interaction is non-local as it involves the total spin of all sites of the electron gas (after the removal of an electron at site $0$).

It would be of interest to identify similar new interaction terms in the low-temperature behavior of a correlated Kondo state with a large band width $D$. Our numerical diagonalization method of Ref.~\onlinecite{rotter08} is limited to a rather small band width, e.g., $N=11$ levels for the results shown in Figs.~\ref{fig:3} and \ref{fig:4}. For such small band widths, the bare and renormalized strong-coupling limits essentially coincide, and we could use our numerical diagonalization method to validate the analytical derivations. For a numerical solution at larger bandwidths, a numerical renormalization group (NRG) technique\cite{wilson75} might be useful. The challenge for NRG is the inclusion of non-local correlations induced by the exchange interaction in the universal Hamiltonian. With numerical solutions at hand it would be interesting to investigate whether the renormalization of the large band width Kondo problem induces any other ``leading irrelevant'' interaction terms around the strong coupling fixed point beyond those we have identified here.

\acknowledgments
We thank S.~Adam, H.~Baranger, L.~Glazman, D.~Goldhaber-Gordon,
R.~Kaul, K.~Le Hur, G.~Murthy, S.~Schmidt, A.~D.~Stone, H.~E.~T\"ureci, and J.~von Delft for helpful discussions. This work was supported in part by the Max-Kade Foundation, the W.~M.~Keck Foundation, U.S. DOE Grant No.~DE-FG-0291-ER-40608 and NSF Grant No.~DMR 0408636.
\renewcommand{\theequation}{A-\arabic{equation}}
\setcounter{equation}{0}

\section*{APPENDIX A}

In this appendix we provide various expressions that are useful
for deriving the effective Hamiltonian. The notation used here
follows the convention of Ref.~\onlinecite{messiah61}.

The product of a hopping operator $\sum_\sigma \,\hat{c}^\dagger_{1,\sigma} \hat{c}_{0,\sigma}$  between sites $0$ and $1$ and its hermitean conjugate can be expressed in terms of the occupation number ($\hat n_0,\, \hat n_1$) and spin ($\hat{\mathbf s}_0, \hat{\mathbf s}_1$) operators at these sites:
\be\label{eq:a1'}
\sum_{\sigma, \sigma'} \,\hat{c}^\dagger_{0,\sigma}
\hat{c}^{\phantom{\dagger}}_{1,\sigma} \hat{c}^\dagger_{1,\sigma'}
\hat{c}^{\phantom{\dagger}}_{0,\sigma'} = \hat n_0 (1-\hat n_1/2) - 2 \hat{\mathbf s}_0\cdot\hat{\mathbf s}_1 \;,
\ee
\be\label{eq:a2'}
\sum_{\sigma, \sigma'} \,\hat{c}^\dagger_{1,\sigma}
\hat{c}^{\phantom{\dagger}}_{0,\sigma} \hat{c}^\dagger_{0,\sigma'}
\hat{c}^{\phantom{\dagger}}_{1,\sigma'} = \hat n_1 (1-\hat n_0/2) - 2 \hat{\mathbf s}_0\cdot\hat{\mathbf s}_1 \;.
\ee

The spin raising and lowering operators,
$\hat{S}_{\pm}\equiv\hat{S}_x\pm i\hat{S}_y$, satisfy, together with $S_z$, the usual $su(2)$ commutation relations
\begin{eqnarray}\label{eq:a10a}
\protect [\hat{S}_{z},\hat{S}_{\pm}]  =  \pm\hat{S}_{\pm}\,,\quad
\protect [\hat{S}_{+},\hat{S}_{-}] =  2\hat{S}_{z}\,
\end{eqnarray}
 while ${\hat{\mathbf S}}^2 = \hat{S}^2_z+\frac{1}{2}(\hat{S}_+\hat{S}_-+\hat{S}_-\hat{S}_+)$. The $su(2)$ commutation relations for the cartesian components of the spin, $\left[\hat{S}_i,\hat{S}_j\right] =i\sum_k\epsilon_{ijk}\hat{S}_k$, can also be written in the form
$\hat{\mathbf S}\times\hat{\mathbf S}=
i\,\hat{\mathbf S}$. The triple scalar product of two spin operators, e.g., ${\mathbf S}'_d$ and ${\mathbf s}_1$ is then given by
\be
\hat{\mathbf S}'_d\cdot(\hat{\mathbf s}_1\times\hat{\mathbf S}'_d)=
i\,\hat{\mathbf s}_1\cdot\hat{\mathbf S}'_d\label{eq:a10}\,.
\ee

Other operator relations between $\hatsb'_d$ and operators on site 1
are
\begin{eqnarray}\label{eq:a7}
\hat{\mathbf S}'^2_d\hat{c}^{\phantom{+}}_{1\pm}\!&=&\!
\hat{c}^{\phantom{+}}_{1\pm}\hat{\mathbf S}'^2_d\!+\!
\frac{3\,\hat{c}^{\phantom{+}}_{1\pm}}{4}
+\hat{c}^{\phantom{+}}_{1\mp}\hat{S}'_{d,\mp}
\mp\hat{c}^{\phantom{+}}_{1\pm}\hat{S}'_{d,z}\,,\quad\quad\\
\hat{\mathbf S}'^2_d\hat{c}^{\dagger}_{1\pm}\!&=&\!
\hat{c}^{\dagger}_{1\pm}\hat{\mathbf S}'^2_d\!+\!
\frac{3\,\hat{c}^{\dagger}_{1\pm}}{4}
+\hat{c}^{\dagger}_{1\mp}\hat{S}'_{d,\pm}
\pm\hat{c}^{\dagger}_{1\pm}\hat{S}'_{d,z}\,.
\label{eq:a7p}
\end{eqnarray}

Useful relations involve the expectation values in the singlet state of observables at site $0$
\be\label{eq:a7'}
\langle \hat{c}^\dagger_{0,\sigma}\hat{c}^{\phantom{\dagger}}_{0,\sigma'}\rangle_s= \langle \hat{c}^{\phantom{\dagger}}_{0,\sigma'}\hat{c}^\dagger_{0,\sigma}\rangle_s= \frac{1}{2}\,\delta_{\sigma \sigma'}\;;\;\;\; \langle \hat n_0\rangle_s =1 \;,
\ee
\begin{eqnarray}\label{eq:a8}
\langle \hat{s}_{0,i}\rangle_s & = & 0 \;,\\\label{eq:a9}
\langle \hat{s}_{0,i}\,\hat{s}_{0,j}\rangle_s &=&\frac{1}{4}\,\delta_{ij}\;,\\ \label{eq:a10'}
\langle \hat{s}_{0,i}\,\hat{s}_{0,j}\,\hat{s}_{0,k} \rangle_s &=&\frac{i}{8}\,\epsilon_{ijk}\;,
\end{eqnarray}
where $\hat{s}_{0,i}$ is the $i$-th cartesian component of $\hat{\mathbf s}_0$ and $\epsilon_{ijk}$ is the third rank antisymmetric tensor.

We can also derive the following expressions for singlet expectation values of the form $\langle \hat H_{sd} \ldots \hat H_{sd} \rangle_s$:
\begin{eqnarray}\label{eq:a6a}
\langle \hat{H}_{sd}\,\hat{n}_0\,\hat{H}_{ds}\rangle_s&=&|t_{0}|^2\, \,\hat{n}_1 \;,\\
\langle \hat{H}_{sd}\,\hat{n}_1\,\hat{H}_{ds}\rangle_s &=&|t_{0}|^2 \;,
\label{eq:a6b}\\
\langle \hat{H}_{sd}\,\hat{n}^2_0\,\hat{H}_{ds}\rangle_s &=&2 \,|t_{0}|^2\,\hat{n}_1 \;,\label{eq:a6c}\\
\langle\hat{H}_{sd}\,\hat{n}^2_1\,\hat{H}_{ds}\rangle_s&=&|t_{0}|^2\,(1+
\hat{n}_1-2\,\hat{n}_{1+}\hat{n}_{1-})\;,\label{eq:a6d}\quad\quad\quad
\end{eqnarray}
and
\begin{eqnarray}
\langle\hat{H}_{sd}\,\hat{\mathbf S}'^2_d\,\hat{H}_{ds}\rangle_s\!&=&\!|t_{0}|^2
\,\left(\frac{3}{4}+\hat{\mathbf S}'^2_d-2\,\hat{\mathbf S}_1
\cdot\hat{\mathbf S}'_d\right)\, \label{eq:a7a}\,,\quad\quad\quad\\
\langle\hat{H}_{sd}\,\hat{\mathbf S}'^4_d\,\hat{H}_{ds}\rangle_s\!&=&\!|t_{0}|^2
\left(\hat{\mathbf S}'^4_d-4\,\hat{\mathbf s}_1\cdot
\hat{\mathbf S}'_d\,\,
\hat{\mathbf S}'^2_d \right. \nonumber \\
&&\left. -\hat{\mathbf s}_1\cdot\hat{\mathbf S}'_d
+\frac{5}{2}\,\hat{\mathbf S}'^2_d+\frac{9}{16}\right)\;.
\label{eq:a7ap}
\end{eqnarray}

Matrix elements of various observables within and between the singlet and triplet manifolds are listed in Tables I, II and III.

\begin{table}[!h]
\begin{tabular}{|c||c|c|c|c|}
\hline
$\langle\psi_1|\hat{O}_1|\psi_2\rangle$
&  $|0,0\rangle$  &
$|1,-1\rangle$  & $|1,0\rangle$   &
$|1,+1\rangle$  \\\hline\hline
 $\langle 0,0|$  & 1 & $\sqrt{2}\,\hat{S}_{1,-}$ & $2\,\hat{S}_{1,z}$  & $-\sqrt{2}\,\hat{S}_{1,+}$ \\
\hline
 $\langle 1,-1|$  & $\sqrt{2}\,\hat{S}_{1,+}$ & $1+2\,\hat{S}_{1,z}$ & $-\sqrt{2}\,\hat{S}_{1,+}$  & 0\\
\hline
 $\langle 1,0|$  & $2\,\hat{S}_{1,z}$ & $-\sqrt{2}\,\hat{S}_{1,-}$ & $1$ & $-\sqrt{2}\,\hat{S}_{1,+}$ \\
\hline
 $\langle 1,+1|$  & $-\sqrt{2}\,\hat{S}_{1,-}$ & $0$ & $-\sqrt{2}\,\hat{S}_{1,-}$  & $1-2\,\hat{S}_{1,z}$\\
\hline
\end{tabular}
\caption{Matrix elements $\langle\psi_1|\hat{O}_1|\psi_2\rangle$
of the operator $\hat{O}_1= \hat H^{(0,1)}_{\rm hop} \hat
H^{(0,1)}_{\rm hop}/|t_0|^2$. The corresponding states $\psi_1$ ($\psi_2$) are listed in the left column (top row), and are
characterized by the quantum numbers $S_{K0},S_{K0,z}$.
}\label{tab:1}
\end{table}

\begin{table}[!h]
\begin{tabular}{|c||c|c|c|c|}
\hline
$\langle\psi_1|\hat{\mathbf s}_0
\cdot\hat{\mathbf S}'_d|\psi_2\rangle$ &
$|0,0\rangle$  &
$|1,-1\rangle$  & $|1,0\rangle$   &
$|1,+1\rangle$  \\\hline\hline
 $\langle 0,0|$  & 0 & $-\hat{S}'_{d,-}/\sqrt{2}$ & $-\hat{S}'_{d,z}$  & $\hat{S}'_{d,+}/\sqrt{2}$ \\
\hline
 $\langle 1,-1|$  & $-\hat{S}'_{d,+}/\sqrt{2}$ & $-\hat{S}'_{d,z}$ & $\hat{S}'_{d,+}/\sqrt{2}$  & 0\\
\hline
 $\langle 1,0|$  & $-\hat{S}'_{d,z}$ & $\hat{S}'_{d,-}/\sqrt{2}$ & 0 & $\hat{S}'_{d,+}/\sqrt{2}$ \\
\hline
 $\langle 1,+1|$  & $\hat{S}'_{d,-}/\sqrt{2}$ & $0$ & $\hat{S}'_{d,-}/\sqrt{2}$  & $+\hat{S}'_{d,z}$\\
\hline
\end{tabular}
\caption{Matrix elements $\langle\psi_1|\hat{\mathbf s}_0\cdot\hat{\mathbf S}'_d|\psi_2\rangle$. Notation as in Table \ref{tab:1}.}\label{tab:2}
\end{table}

\begin{table}[!h]
\begin{tabular}{|c||c|}
\hline
$\langle\psi_1|\hat{O}_2|\psi_2\rangle$ &
$|0,0\rangle$  \\\hline\hline
$\langle 1,-1|$  & $\phantom{\Big{)}}
\frac{1}{\sqrt{2}}(-\hat{c}_{1-}^{\phantom{\dagger}}
\hat{\mathbf S}'^2_d\hat{c}_{1+}^{\dagger}+\hat{c}_{1+}^{\dagger}
\hat{\mathbf S}'^2_d\hat{c}_{1-}^{\phantom{\dagger}})\phantom{\Big{)}}$ \\
\hline
$\langle 1,0|$  & $\genfrac{}{}{0pt}{0}{\phantom{\Big{)}}
\frac{1}{2}(-\hat{c}_{1+}^{\phantom{\dagger}}
\hat{\mathbf S}'^2_d\hat{c}_{1+}^{\dagger}+\hat{c}_{1-}^{\phantom{\dagger}}
\hat{\mathbf S}'^2_d\hat{c}_{1-}^{\dagger}\phantom{\Big{)}}}
{\phantom{\Big{)}}
-\hat{c}_{1-}^{\dagger}\hat{\mathbf S}'^2_d\hat{c}_{1-}^{\phantom{\dagger}}
+\hat{c}_{1+}^{\dagger}\hat{\mathbf S}'^2_d\hat{c}_{1+}^{\phantom{\dagger}})\phantom{\Big{)}}}$ \\
\hline
 $\langle 1,+1|$  & $\phantom{\Big{)}}
\frac{1}{\sqrt{2}}(+\hat{c}_{1+}^{\phantom{\dagger}}
\hat{\mathbf S}'^2_d\hat{c}_{1-}^{\dagger}-\hat{c}_{1-}^{\dagger}
\hat{\mathbf S}'^2_d\hat{c}_{1+}^{\phantom{\dagger}})\phantom{\big{)}}$ \\
\hline
\end{tabular}
\caption{Matrix elements $\langle\psi_1|\hat{O}_2|\psi_2\rangle$
of the operator $\hat{O}_2=\hat H^{(0,1)}_{\rm hop}\hat{\mathbf S}'^2_d
\hat H^{(0,1)}_{\rm hop}/|t_0|^2$.  Notation as in Table \ref{tab:1}.}\label{tab:3}
\end{table}

\end{document}